\documentclass[aps,prl,preprint,superscriptaddress]{revtex4-1}
\usepackage{amsmath,amssymb,graphicx}
\usepackage{color,epsfig,rotate}

\newcommand{\Ag}{AgVOAsO$_{4}$\,}
\newcommand{\Ba}{Ba$_{3}$Mn$_{2}$O$_{8}$\,}

\begin{document}


\title{Field-induced double dome and Bose-Einstein condensation in the crossing quantum spin chain system \Ag}


\author{Franziska Weickert}
\email[]{weickert@lanl.gov}
\affiliation{NHMFL, Florida State University, Tallahassee, FL 32310, USA}
\author{Adam A. Aczel}
\email[]{aczelaa@ornl.gov}
\affiliation{Neutron Scattering Division, Oak Ridge National Laboratory, Oak Ridge, TN 37831, USA}
\author{Matthew B. Stone}
\affiliation{Neutron Scattering Division, Oak Ridge National Laboratory, Oak Ridge, TN 37831, USA}
\author{V. Ovidiu Garlea}
\affiliation{Neutron Scattering Division, Oak Ridge National Laboratory, Oak Ridge, TN 37831, USA}
\author{Chao Dong}
\affiliation{ISSP, International MegaGauss Science Laboratory, University of Tokyo, Kashiwa Chiba 277-8581, Japan}
\author{Yoshimitsu Kohama}
\affiliation{ISSP, International MegaGauss Science Laboratory, University of Tokyo, Kashiwa Chiba 277-8581, Japan}
\author{Roman Movshovich}
\affiliation{MPA-CMMS, Los Alamos National Laboratory, Los Alamos, NM 87545 USA}
\author{Albin Demuer}
\affiliation{GHMFL, CNRS, 38042 Grenoble cedex 9, France}
\author{Neil Harrison}
\affiliation{MPA-Mag, Los Alamos National Laboratory, Los Alamos, NM 87545, USA}
\author{Monika B. Gam\.{z}a}
\affiliation{MPI CPfS, N\"{o}thnitzer Str. 40, 01187 Dresden, Germany}
\affiliation{Jeremiah Horrocks Institute for Mathematics, Physics, and Astrophysics, University of Central Lancashire, Preston PR1 2HE, UK}
\author{Alexander Steppke}
\affiliation{MPI CPfS, N\"{o}thnitzer Str. 40, 01187 Dresden, Germany}
\author{Manuel Brando}
\affiliation{MPI CPfS, N\"{o}thnitzer Str. 40, 01187 Dresden, Germany}
\author{Helge Rosner}
\affiliation{MPI CPfS, N\"{o}thnitzer Str. 40, 01187 Dresden, Germany}
\author{Alexander A. Tsirlin}
\affiliation{MPI CPfS, N\"{o}thnitzer Str. 40, 01187 Dresden, Germany}
\affiliation{Experimental Physics VI, Augsburg University, 86135 Augsburg, Germany}


\date{\today}

\begin{abstract}
We present inelastic neutron scattering data on the quantum paramagnet AgVOAsO$_4$ that establish the system is a $S=1/2$ alternating spin chain compound and provide a direct measurement of the spin gap. We also present experimental evidence for two different types of field-induced magnetic order between $\mu_{0}H_{c1}$~$=$~8.4~T and $\mu_{0}H_{c2}$~$=$~48.9~T, which may be related to Bose-Einstein condensation (BEC) of triplons. Thermodynamic measurements in magnetic fields up to 60~T and temperatures down to 0.1~K reveal a $H-T$ phase diagram consisting of a dome encapsulating two ordered phases with maximum ordering temperatures of 3.8~K and 5.3~K respectively. This complex phase diagram is not expected for a single-$\vec{Q}$ BEC system and therefore establishes AgVOAsO$_4$ as a promising multi-$\vec{Q}$ BEC candidate capable of hosting exotic vortex phases. 
\end{abstract}


\maketitle


The investigation of Bose-Einstein condensation (BEC) in quantum paramagnets\cite{giamarchi_08} has been a fruitful area of research over the last two decades. BEC has been identified in more than 20 different compounds based on magnetic moments of Cu, V, Cr, Ni or Mn\cite{zapf_14,nikuni_00,jaime_04,zapf_06,garlea_07,samulon_08,aczel_09,aczel_09_2}. The materials studied often consist of interacting $S$~$=$~1/2 spin dimers. In this case, a non-magnetic $S$~$=$~0 singlet ground state is separated from the excited $S$~$=$~1 triplet states by a spin gap $\Delta$, with a finite dispersion for the triplet excitation arising from the interdimer exchange interactions. An applied magnetic field $H$ splits the triplet into its three branches according to their $S^{z}$ quantum number.  As the magnetic field increases, the spin gap closes at a critical field $H_{c1}$ and generates field-induced magnetic order. If $O(2)$ rotational invariance is preserved above $H_{c1}$, then this ordered state is equivalent to BEC of $S^{z}$~$=$~1 triplons\cite{matsubara_56,batyev_84,affleck_91,giamarchi_99,giamarchi_08}.

The simplest version of a field-induced triplon BEC arises from closing the spin gap in a quantum paramagnet with a triplet dispersion characterized by a single minimum in the Brillouin zone. The bosons condense into a single-$\vec{Q}$ BEC state best described by XY antiferromagnetic order with a commensurate magnetic structure above $H_{c1}$ and a saturated paramagnet is generated above $H_{c2}$. On the other hand, in quantum paramagnets with significant frustration due to competing interdimer exchange interactions, the triplet dispersion may be modified so that there are several degenerate minima in a single Brillouin zone. The $H-T$ phase diagram of these multi-$\vec{Q}$ BEC systems has been predicted to be extremely rich with multiple field-induced phases expected in a given material and the tantalizing possibility of realizing topological spin textures, such as magnetic vortex crystals\cite{kamiya_14}. Magnetic vortices are close relatives of the magnetic skyrmions observed in metallic MnSi\cite{muehlbauer_09} and in the Mott insulator Cu$_{2}$OSeO$_{3}$\cite{seki_12}. 

One promising material recently discussed in the context of multi-$\vec{Q}$ BEC is \Ba, with inelastic neutron scattering (INS) work\cite{stone_08a, stone_08b} reporting evidence for frustrated interdimer exchange and a triplet dispersion consisting of several energy minima in a single Brillouin zone. Intriguingly, thermodynamic and torque magnetometry measurements determined that the $H$-$T$ phase diagram was more complex than expected for a single-$\vec{Q}$ BEC system, with two ordered phases I and II found for all field orientations except $\vec{H}$~$\parallel$~$c$ where only one ordered phase was uncovered\cite{samulon_08, samulon_10}. Subsequent neutron diffraction measurements in a horizontal scattering plane with an applied field $\vec{H}$~$\parallel$~$a^*$ identified phases I and II as an incommensurate spin spiral and spin density wave state respectively\cite{stone_15}, but nuclear magnetic resonance (NMR) work argued that the BEC description only holds for $\vec{H}$~$\parallel$~$c$\cite{suh_11}. Other material candidates are therefore required, if one hopes to identify exotic multi-$\vec{Q}$ BEC states, including the magnetic vortex crystal, in the laboratory. 

\begin{figure}
	\includegraphics[width=3.3in]{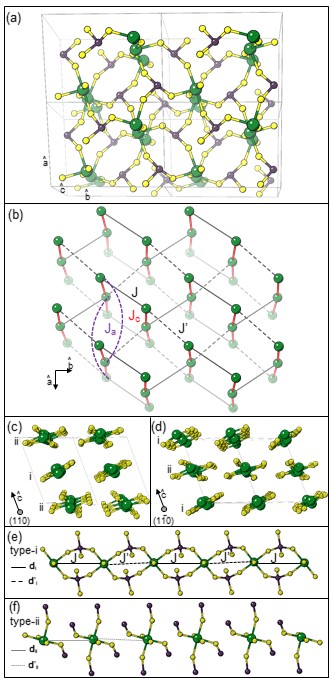}
	\caption{\label{FigX} Crystal structure of \Ag. V atoms are large green spheres, O atoms are small yellow spheres and As atoms are black spheres. The Ag atoms are not shown for clarity. (a) Crystal structure viewed obliquely along the $c$-axis. (b) Exchange paths viewed roughly along the $c$-axis. For clarity, the $J_{a}$ interaction is shown only for a single $ac$-plane and only the V sites are shown. (c) The crystal structure viewed along the (110) direction. (d) The crystal structure viewed along the (1$\overline{1}$0) direction.  In (c) and (d), chains of type-i and type-ii are labeled accordingly. (e) Type-i structural chain with exchange interactions $J$ and $J'$ labeled for vectors $d_{i}$ and $d'_{i}$ respectively. (f) Type-ii structural chain with vectors $d_{ii}$ and $d'_{ii}$ shown.}
\end{figure} 

To this end, the V$^{4+}$, $S$~$=$~1/2 compound \Ag is a promising multi-$\vec{Q}$ BEC candidate. The monoclinic crystal structure consists of corner-sharing VO$_{6}$ octahedra that form structural chains along the crystallographic $c$-direction; these chains are linked to one another via AsO$_4$ tetrahedra to form a three-dimensional network. The crystal structure is shown in Fig.\ref{FigX}. Figure \ref{FigX} (a) illustrates the crystal structure viewed roughly along the $c$-axis. There are chain structures along this axis with a V-V distance of 3.639\AA\ at T = 20 K. In Fig.\ref{FigX} (b) we reproduce the proposed exchange paths\cite{tsirlin_11} showing only the V sites. We note that the $J_{a}$ and $J_{c}$ exchange connect spins between layers. In Figures \ref{FigX}(c) and (d) we show views along the (110) and (1$\overline{1}$0) direction.  There are two types of structural chains along these directions which we label as type-i and type-ii.  The type-i chain is the previously proposed magnetic alternating chain structure.  The type-ii chains are nearly orthogonal to the type-i chains. The two chains have very similar vectors and distances between V sites, but their bonding is quite different. The type-i alternating chain structure has a much more planar configuration of coordinated oxygen atoms than the type-ii chain. For example, the middle layer of chains in Fig.\ref{FigX}(c) shows this planar nature. When viewed along the (1$\overline{1}$0) direction, it is the top and bottom layers of chains, as shown in Fig.\ref{FigX}(d), that have the more planar oxygen coordination. Figures \ref{FigX}(e) and (f) illustrate the type-i and type-ii chain structures. The type-i chain shown in Fig. \ref{FigX}(e) is the one that has a more planar coordination of oxygen atoms and has been considered to be the magnetic alternating chain based upon DFT calculations\cite{tsirlin_11}. $d_{i}$ is considered to be the strong or dimer-bond along the alternating chain, $J$, and $d'_{i}$ is the inter-dimer bond, $J'$. The dimer-dimer vector is $u_{0}=d_{i}+d'_{i}$, which corresponds to the (110) and (1$\overline{1}$0) directions depending upon which ab-plane the chain resides within. The V-V vectors at 20K are $d_{i}$= [0.488, -0.525, 0.060] and [0.488, 0.525, 0.060] ($|d_{i}| = 5.59$\AA), $d'_{i}$ =  [0.512, -0.475,  -0.060] and [0.512, 0.475, -0.060] ($|d'_{i}| = 5.56$\AA), $d_{ii}$ =  [0.512, 0.525, -0.060] and [-0.512, 0.525, 0.060] ($| d'_{ii}| = 5.90$ \AA), $d’_{ii}$ =  [0.488, 0.475, 0.060]  and [-0.488, 0.475, -0.060] ($|d'_{ii}| = 5.23$\AA).

Bulk characterization and $^{75}$As NMR measurements propose \Ag\ to be a quantum paramagnet based on alternating spin chains \cite{xu_00, stone_14} with a spin gap $\Delta$~$=$~1.1~meV, $\mu_{0}H_{c1}$~$=$~10~T, a saturation field $\mu_{0}H_{c2}$~$=$~48.5~T, and an intrachain exchange ratio $\alpha$~$=$~$J'/J$~$\simeq$~0.6-0.7\cite{tsirlin_11, ahmed_17}. DFT calculations furthermore predict significant, competing interchain exchange interactions leading to a large degree of magnetic frustration, which may produce the complicated triplet dispersion that is a prerequisite for multi-$\vec{Q}$ BEC. The relatively small spin gap ensures that these field-induced ordered states are accessible in the laboratory. 
In this work, we first show INS results on polycrystalline \Ag (see Ref.~\cite{tsirlin_11} for synthesis details) that establish the alternating spin chain model \cite{tsirlin_11} and provide a direct measurement of the spin gap. Second, we present a comprehensive study of the field-induced magnetic order in this material. Our combined specific heat, magnetization, and magnetocaloric effect (MCE) measurements on polycrystalline samples establish a complex $H-T$ phase diagram with two different field-induced ordered states, which is not expected for a single-$\vec{Q}$ BEC system. Therefore, \Ag is a strong candidate for hosting multi-$\vec{Q}$ BEC.

\begin{figure}
\includegraphics[width=3.3in]{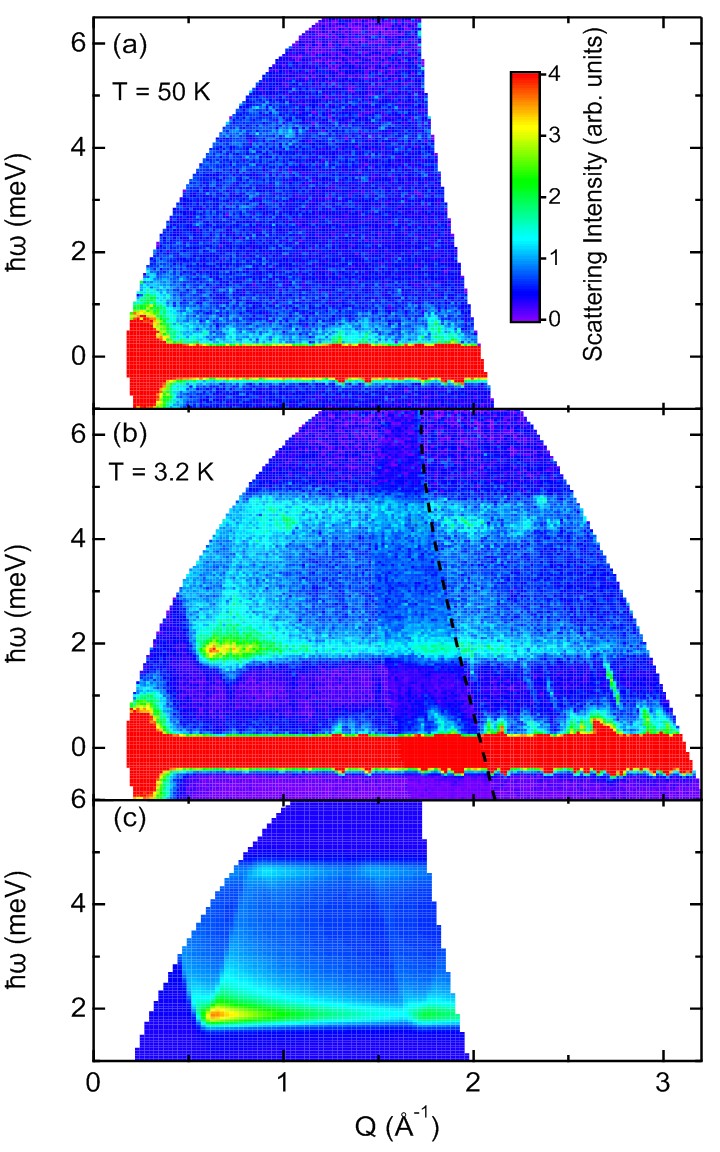}
\caption{\label{Fig1} (a), (b) Color contour plots of the $E_i$~$=$~7.5~meV HYSPEC data for AgVOAsO$_4$ at $T$~$=$~50~K and $T$~$=$~3.2~K. A gapped magnetic excitation spectrum is clearly visible in the low-temperature data. The data to the left of the black dashed line in (b) was collected using a single orientation for the HYSPEC detector bank. (c) Color contour plot for the alternating chain model described in the text with best fit parameters $J$~$=$~3.43(7)~meV and $J'$~$=$~2.25(9)~meV. The INS data was only fit over the $Q$-$\hbar \omega$ region presented in this figure.}
\end{figure} 

Inelastic neutron scattering data were collected on 10~g of polycrystalline \Ag using the HYSPEC spectrometer of the Spallation Neutron Source, Oak Ridge National Lab (ORNL). All data were collected using incident energies of $E_i$~$=$~7.5 or 15 meV, with corresponding Fermi chopper frequencies of 180 and 300~Hz, resulting in instrumental energy resolutions of 0.3 and 0.7 meV (Gaussian full-width half-maximum [FWHM]) respectively at the elastic line. The HYSPEC instrument is able to extend the range of measured wavevector transfer, $Q$, at a fixed incident energy by moving its detector bank to larger scattering angles. This was done for both the $E_i$~$=$~7.5 and 15 meV measurements presented here. A liquid He cryostat was used during the measurements to achieve temperatures of 3.2 - 200~K. 

Color contour plots of the finer energy resolution $E_i$~$=$~7.5~meV HYSPEC data at $T$~$=$~50 and 3.2 K are presented in Fig.~\ref{Fig1}(a) and (b) respectively. The lower $T$ data shows a band of scattering between 1.5 and 5 meV energy transfer, $\hbar\omega$, with intensity that decreases rapidly as a function of wavevector transfer, $Q$. The intensity of this excitation also displays a distinct, oscillatory $Q$-dependence. Both the temperature and wavevector dependence of this mode are hallmarks of magnetic fluctuations arising from excited triplet states in a dimerized magnet.  No higher energy magnetic excitations were observed in the coarser energy resolution $E_i$~$=$~15~meV HYSPEC data.  

\begin{figure}
\includegraphics[width=3.3in]{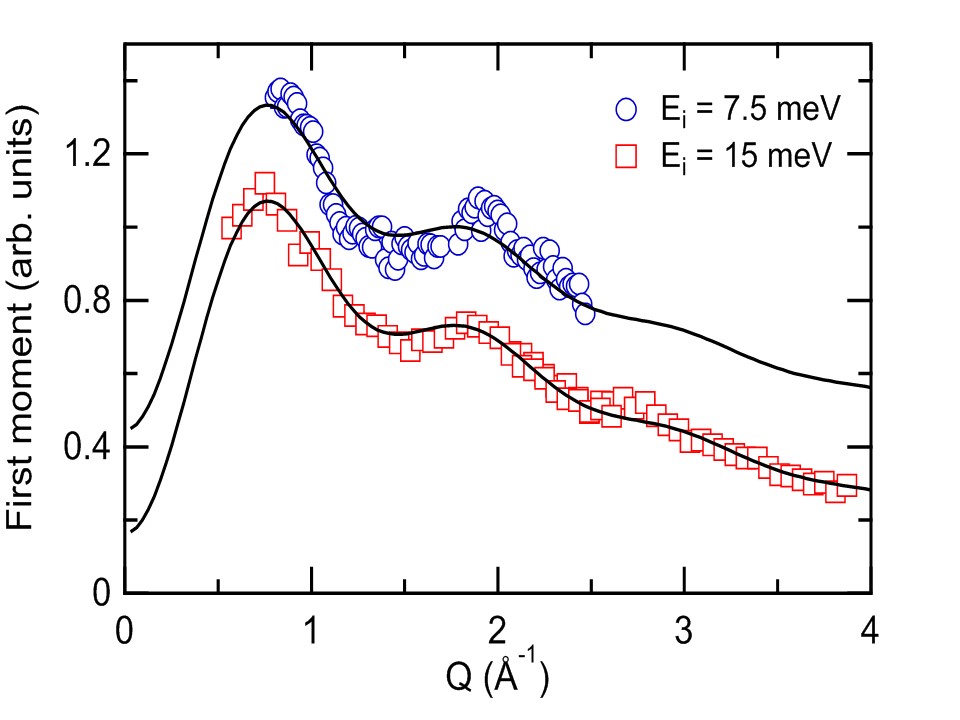}
\caption{\label{Fig2} The powder-averaged first frequency moment as a function of wavevector transfer for both incident energies used in the HYSPEC measurement. An overall normalization factor has been applied so the two datasets can be plotted with the same y-axis. The solid curves represent a simultaneous fit of the data to the first frequency moment expression described in the text with a single exchange interaction.}
\end{figure}

With the magnetic origin of the spectrum established, we next performed an analysis of the INS data using the powder-averaged first frequency moment $\langle E(Q) \rangle$ approach. More specifically, we turn to the following equation, valid for an isotropic spin system with Heisenberg exchange interactions \cite{hohenberg_74, stone_08b, tassel_10}: 
\begin{equation}
\left\langle E(Q)\right\rangle \propto -\Sigma J_{j} \left\langle S_{0}\cdot S_{d_{j}}\right\rangle |f(Q)|^{2} \left(1-\dfrac{\sin Q d_{j}}{Q d_{j}}\right)
\end{equation} 
where $f(Q)$ is the magnetic form factor for V$^{4+}$, $J_{j}$ is the exchange interaction between magnetic ions with spin $S$ separated by a distance $d_j$, and $\left\langle S_{0}\cdot S_{d_{j}}\right\rangle$ is the two-spin correlation function for this pair. The integration ranges used to extract the first frequency moment from the $E_i$~$=$~7.5 and 15 meV measurements were set to 1.5 - 5 meV and 1.25 - 5 meV respectively. These data are shown in Fig.~\ref{Fig2}.  Both measurements have an oscilatory intensity that decreases as $Q$ increases.  The two datasets were simultaneously fit to Eq.~(1) considering only a single value of $d_{j} = d$, but with a multiplicative prefactor and a constant background unique to each value of incident energy. The results of this comparison are shown as solid curves in Fig.~\ref{Fig2}. This analysis is able to reproduce the oscillatory wavevector-dependence of the first moment with $d=5.63(4)$~\AA. A closer look at the \Ag crystal structure (Fig.\ref{FigX}) reveals two possible V$^{4+}$ alternating chains, both confined to the $ab$-plane, that may be consistent with this $d$ value. The first alternating chain candidate (type i) has distances of $d_{i}=$ 5.59\AA\ and $d'_{i}=$5.56~\AA~between spins at 20~K, while the second chain candidate (type ii) has distances of $d_{ii}=$5.90\AA\ and $d'_{ii}=$5.23~\AA. Band structure calculations determined that the magnetic properties of \Ag arise from type-i chains, with the distance $d_{i}$~$=$~5.59~\AA~corresponding to the strong, dimer bond in the system\cite{tsirlin_11}. The determined value of $d$ from the first moment analysis agrees well with this designation.

We proceeded to calculate the dynamical structure factor $S(Q,\omega)$ for \Ag using the expression for the alternating Heisenberg spin chain model:
\begin{equation}
S(\vec{Q},\omega) = A |f(Q)|^{2} [1-cos(\vec{Q} \cdot \vec{d})] \delta[\hbar\omega-\hbar\omega(\vec{Q})]
\end{equation}
where $A$ is a multiplicative prefactor and $\vec{d}$ is the vector between spins of the dimer pair (distance $d_{i}$). The first order approximation to the alternating chain model dispersion $\hbar \omega(\vec{Q})$ is given by:
\begin{equation}
\hbar\omega(\vec{Q}) = J - \frac{J'}{2} cos(\vec{Q} \cdot \vec{u_{0}})
\end{equation}
where $J$ and $J'$ are the exchange interactions of the alternating chain and $\vec{u_{0}}$ is the vector connecting the centers of two adjacent dimers.  Prior bulk characterization measurements have established that $\alpha=J'/J\approx 0.65$ for \Ag. This larger value of $\alpha$ places the potential dispersion for \Ag far from the first order approximation for the alternating chain model, so we used a modified version of Eq.~(3) in our modeling with terms up to third order in $\alpha$ as described in Refs.~\cite{brooks1973} and \cite{barnes1999}. To facilitate a direct comparison with our INS data, we powder-averaged Eq.~(2) according to the following:
\begin{equation}
S(Q, \omega) = \int \frac{d\Omega_{Q}}{4 \pi} S(\vec{Q}, \omega)
\end{equation}
More specifically, we calculate $S(\vec{Q}, \omega)$ over spherical shells in $Q$ space at fixed values of energy transfer with $\vec{d}$ and $\vec{u_{0}}$ set to the 20~K crystal structure values\cite{tsirlin_11} for the proposed $[1\bar{1}0]$ and $[110]$ alternating chains.  We account for the different chain directions in adjacent $ab$-planes by including equal contributions from these two chain types in our model. The modified spectrum was then multiplied by $|f(Q)|^2$ and convolved with a Gaussian approximation for the instrumental energy and wavevector resolution. A constant background and the multiplicative prefactor were incorporated as fitting parameters of the calculated spectrum in comparison to the measured data. This modeling can accurately reproduce the wavevector and energy-dependence of the measurement when $J$~$=$~$3.43(7)$~meV (39.8(7)K) and $J'$~$=$~$2.25(9)$~meV (26.2(1)K), as shown in Fig.~\ref{Fig1}(c). The determined exchange interactions and their ratio $\alpha$~$=$~$J'/J$~$=$~0.66(3) are in good agreement with values determined in previous work\cite{tsirlin_11, ahmed_17}.

\begin{figure}
\includegraphics[width=3.3in]{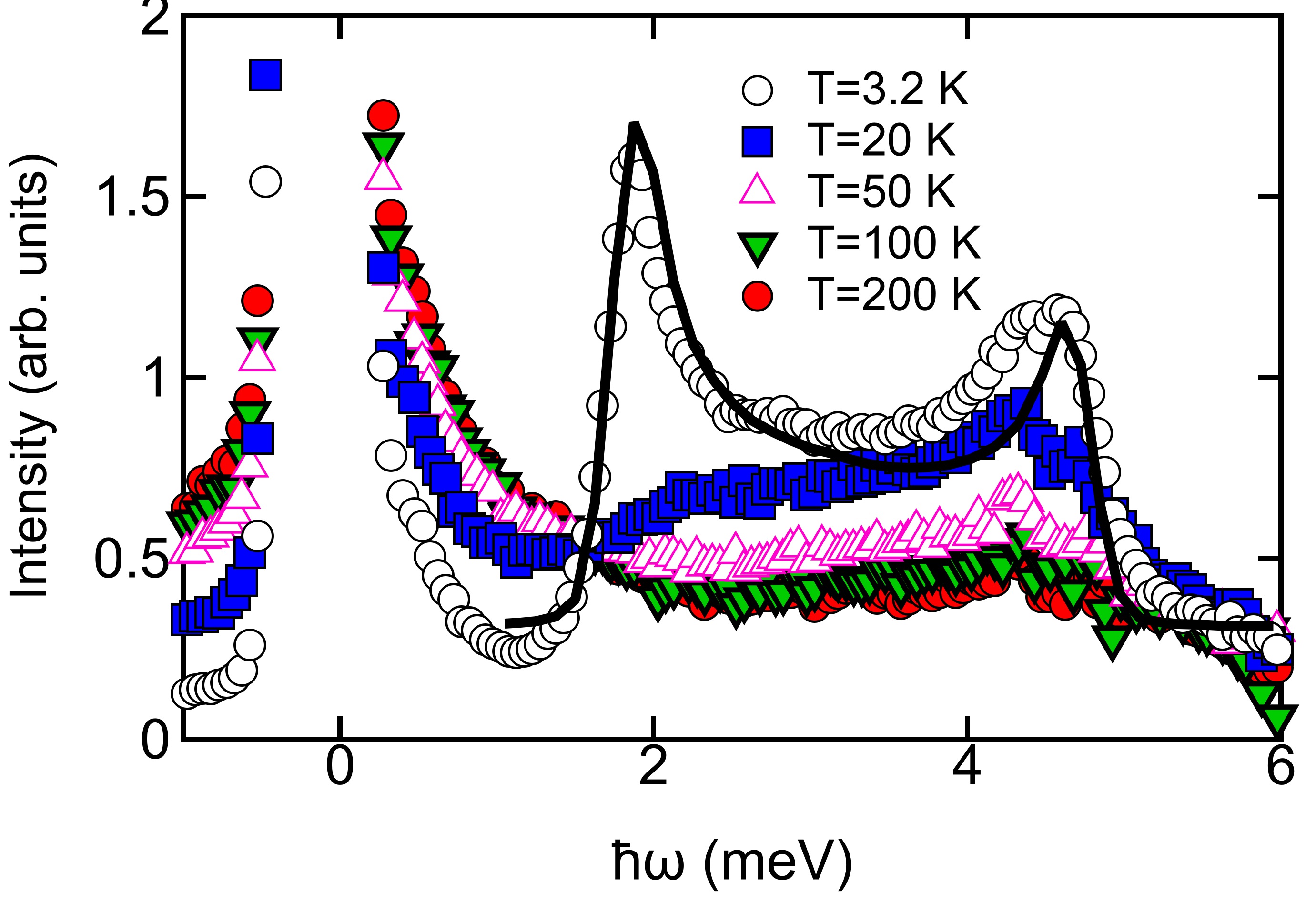}
\caption{\label{Fig3} Temperature-dependence of a constant-$Q$ cut with an integration range $Q$~$=$~[0.35, 2]~\AA$^{-1}$ from the $E_i$~$=$~7.5~meV HYSPEC dataset. The solid curve superimposed on the 3.2~K data represents a constant-$Q$ cut taken from the best fit simulation shown in Fig.~\ref{Fig1}(c). }
\end{figure}  

Figure~\ref{Fig3} shows constant-$Q$ cuts (integration range $Q$~$=$~[0.35, 2]~\AA$^{-1}$) for the $E_i$~$=$~7.5~meV data at different temperatures. We superimpose a cut through our model in the figure and find a good overall agreement with the data, however, 
the measured scattering intensity is not fully-captured near the top of the band between 3 and 5 meV. A large portion of this extra intensity persists up to high temperatures. These combined findings are consistent with a small phonon contribution to the measured spectrum that we do not account for in our modeling. On the other hand, this excess scattering may also arise from two-triplon excitations. This scenario is particularly plausible for quantum paramagnets like \Ag where the triplet excitation bandwidth is much larger than the spin gap, as the continuum of two-triplon modes will then extend down into the single-particle regime \cite{stone_06} that we have modeled above. Single crystal INS data will ultimately be required to definitively establish the origin of the additional scattering in the magnetic excitation spectrum of \Ag. 


\begin{figure}
\includegraphics[width=3.5in]{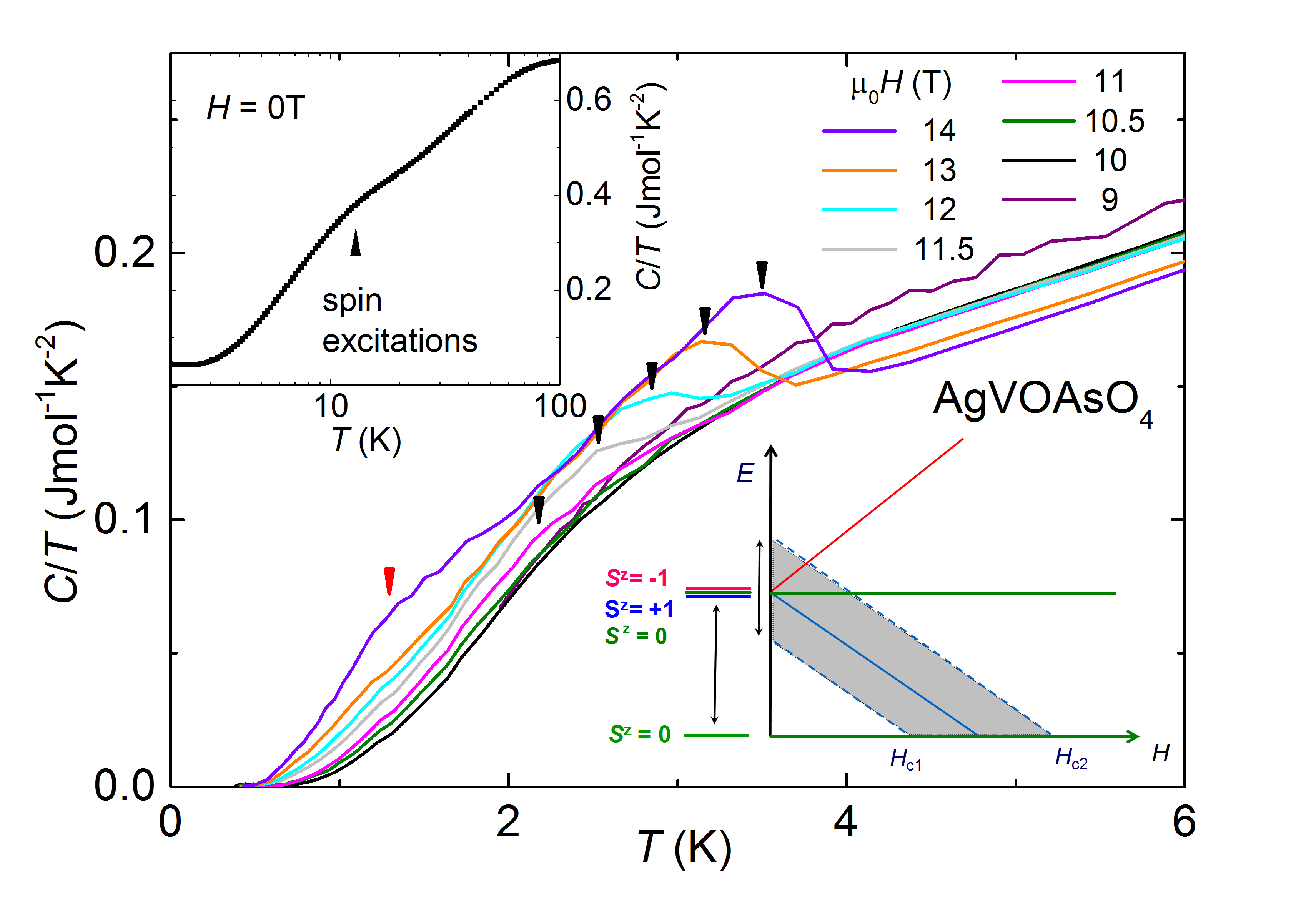}
\caption{\label{Cplow} Specific heat divided by temperature $C(T)/T$ vs $T$ for magnetic fields between 9~T and 14~T with nuclear Schottky contributions subtracted. The measurement at 11~T is the first curve that develops a broad maximum at $\sim$~2.2~K that becomes more pronounced and shifts to higher $T$ in stronger fields (marked with black arrows). Additionally, a second anomaly appears at $\sim$~1~K in the 14~T measurement, as indicated by the red arrow. The upper inset shows $C(T)/T$ in zero magnetic field, with a broad hump around 13~K arising from the thermal population of the triplet state. The lower inset displays the energy level scheme for a spin dimer system with both intradimer and interdimer exchange coupling under the application of an external magnetic field. } 

\end{figure} 

With the alternating chain character confirmed by INS measurements, we now examine the magnetic field and temperature-dependent phase diagram of \Ag. The small spin gap should be closed at an experimentally-accessible critical field $H_{c1}$ and this allows one to search for field-induced magnetic order. We performed specific heat $C(T)$ measurements using a standard relaxation technique with a Quantum Design Physical Property Measurement System in magnetic fields up to 14~T to look for signatures of magnetic order. The upper inset in Fig.~\ref{Cplow} shows the specific heat divided by temperature $C(T)/T$ in zero field, with a broad hump occurring at 13~K as expected for an interacting, alternating spin chain system with a non-magnetic singlet ground state. We put considerable effort into the preparation of a non-magnetic reference compound to subtract the phonon contribution. Unfortunately, the non-magnetic analog AgTiOAsO$_{4}$ does not exist or at least cannot be synthesized under standard conditions.   

\begin{figure}
\includegraphics[width=3.5in]{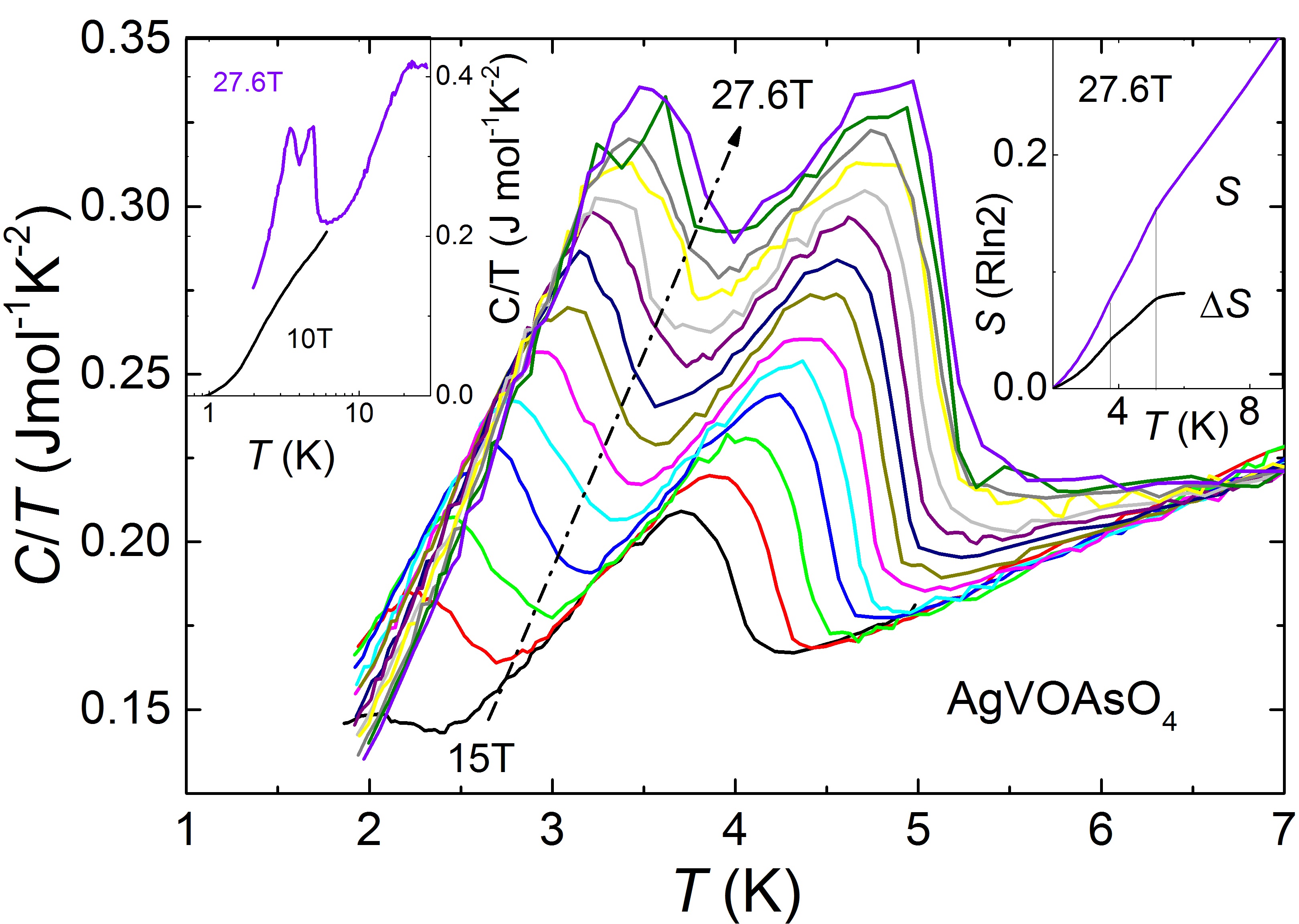}
\caption{\label{Cphigh} Specific heat divided by temperature $C(T)/T$ vs $T$ for magnetic fields between 15~T and 27.6~T. Two distinct maxima are observed that move to higher temperatures with increasing field and represent clear evidence for multiple field-induced phase transitions below the onset of the saturated paramagnetic phase at $H_{c2}$. The left inset displays $C(T)/T$ for the 10~T and maximum field 27.6~T measurements. The right inset shows the calculated entropy $S = \int C(T)/T dT$ in units R$\ln$2 for the 27.6~T measurement with (i.e. $\Delta S$) and without (i.e. $S$) subtraction of the 10~T $C(T)/T$ data. }
\end{figure}

The specific heat divided by temperature $C(T)/T$ vs $T$ for magnetic fields between 9 - 14~T, with appropriate nuclear Schottky contributions subtracted (discussion below), is shown in Fig.~\ref{Cplow}. We observe the onset of a broad maximum at 2.2~K in the 11~T data (indicated by a black arrow), in addition to the hump observed at 13K. This anomaly becomes more pronounced in magnetic fields $\ge$~11.5~T and provides the first evidence for field-induced magnetic order in this material. A close look at the data obtained at 14~T reveals that the first maximum has shifted to 3.7~K and a second feature in the data is now visible at 1~K, as indicated by the red arrow in Fig.~\ref{Cplow}. To map out a larger region of the $H-T$ phase diagram, we extended our measurements to 27.6~T in a resistive magnet at the Laboratoire National des Champs Magnetiques Intenses in Grenoble using a relaxation dual slope technique\cite{riegel_86}. As shown in Fig.~\ref{Cphigh}, we observe that both maxima in $C/T(T)$ develop into distinct $\lambda$-anomalies for $H$~$>$~15~T that are typical for second order phase transitions. As mentioned before, specific heat measurements in high magnetic fields often show a significant nuclear Schottky contribution $[C(T)/T]_{ns}$ at low $T$ caused by isotopes with non-zero nuclear spin. In \Ag , $^{107}$Ag and $^{109}$Ag ($I$~$=$~1/2, 50\% natural abundance each), $^{51}$V (7/2, 99\%), and $^{75}$As (3/2, 100\%) contribute to this effect. We subtract $[C(T)/T]_{ns}$~$=$~$a_{0} T^{-3}$ from the data and estimate a prefactor $a_{0}$= 1.55~mJ-K/mol at low fields that increases by 30\% in the 14~T measurement. Enlarged $a_{0}$ values above $H_{c1}$ are a further indication for field-induced order, as this implies that the internal magnetic field detected by the nuclear spins is strongly enhanced.

The maximum entropy for a spin-1/2 dimer system is R$\ln 2$, which is released when the thermal energy is significantly larger than the intradimer exchange $J$ \cite{brambleby_17}. Since our INS measurements described above found $J$~$=$~40~K, we expect the high-$T$ entropy regime to onset well above the 13~K maximum observed in the 0~T $C(T)/T$ data. In the left inset of Fig.~\ref{Cphigh}, we show $C(T)/T$ for the 10~T and 27.6~T measurements on the same scale to facilitate an easier comparison. The right inset shows the calculated entropy for the 27.6~T measurement with (i.e. $\Delta S$) and without (i.e. $S$) subtraction of the 10~T $C(T)/T$ data. About 30\% of the maximum entropy R$\ln 2$ is recovered upon warming up to 10~K with only a few percent released at each of the two phase transitions. As shown later, these small differences in entropy are sufficiently large to track the phase boundaries in magnetocaloric effect (MCE) measurements under quasiadiabatic conditions.  


\begin{figure}
\includegraphics[width=3.5in]{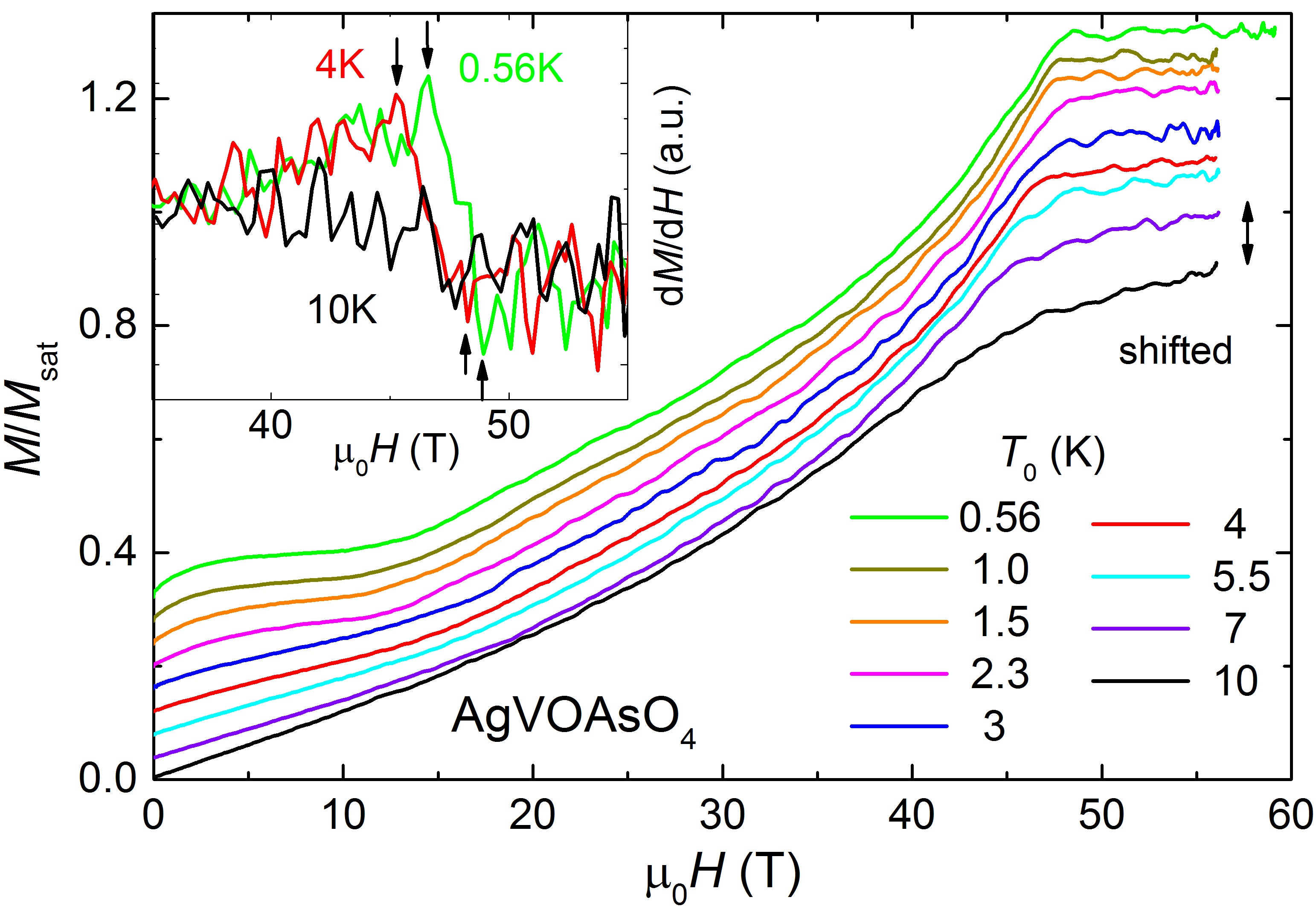}
\caption{\label{MvsH_high} Magnetization $M(H)$ vs magnetic field $H$ measured in high fields to 60~T is shown for initial temperatures $T_{0}$ between 0.56~K and 10~K (data are offset vertically for better visibility). The inset shows $dM/dH$ close to $H_{c2}$ for the measurements at 0.56~K and 4~K where both a sharp maximum and minimum are visible, indicating a double phase transition. No sharp anomalies are observed in the 10~K $dM/dH$ data.}
\end{figure} 

We continue our high-field study with extraction magnetometry experiments measuring $M(H)$ inside a capacitor-driven pulsed magnet at the Pulsed Field Facility of the National High Magnetic Field Laboratory (NHMFL). Signatures in $M(H)$ have been used successfully in the past to estimate the phase diagram of related compounds \cite{sebastian_07}. Fig.~\ref{MvsH_high} shows the magnetization of \Ag\ measured up to 60~T with the initial sample temperature $T_{0}(H=0)$ in the range between 0.56~K and 10~K. At low $T$, we can distinguish three different field regions. In fields up to 5~T, the magnetization is dominated by the paramagnetic behavior of unpaired spins arising from defects in the crystal structure\cite{ahmed_17}. Above 5~T these unpaired spins are fully-polarized, which leads to a constant background in the quantum paramagnetic state. Starting at a critical field $H_{c1}$ near 10~T, the magnetization increases monotonically and reaches saturation at about 48~T. The field-dependence of $M(H)$ can be reproduced with an interacting, alternating spin chain model \cite{tsirlin_11}. The magnetization close to full saturation reveals two distinct changes in the slope at $\mu_{0}H_{c2'}$~$=$~46.6~T and $\mu_{0}H_{c2}$~$=$~48.8~T, as can be seen more clearly in the derivative $\partial M/\partial H$ in the inset of Fig~\ref{MvsH_high}. 

\begin{figure}
\includegraphics[width=3.5in]{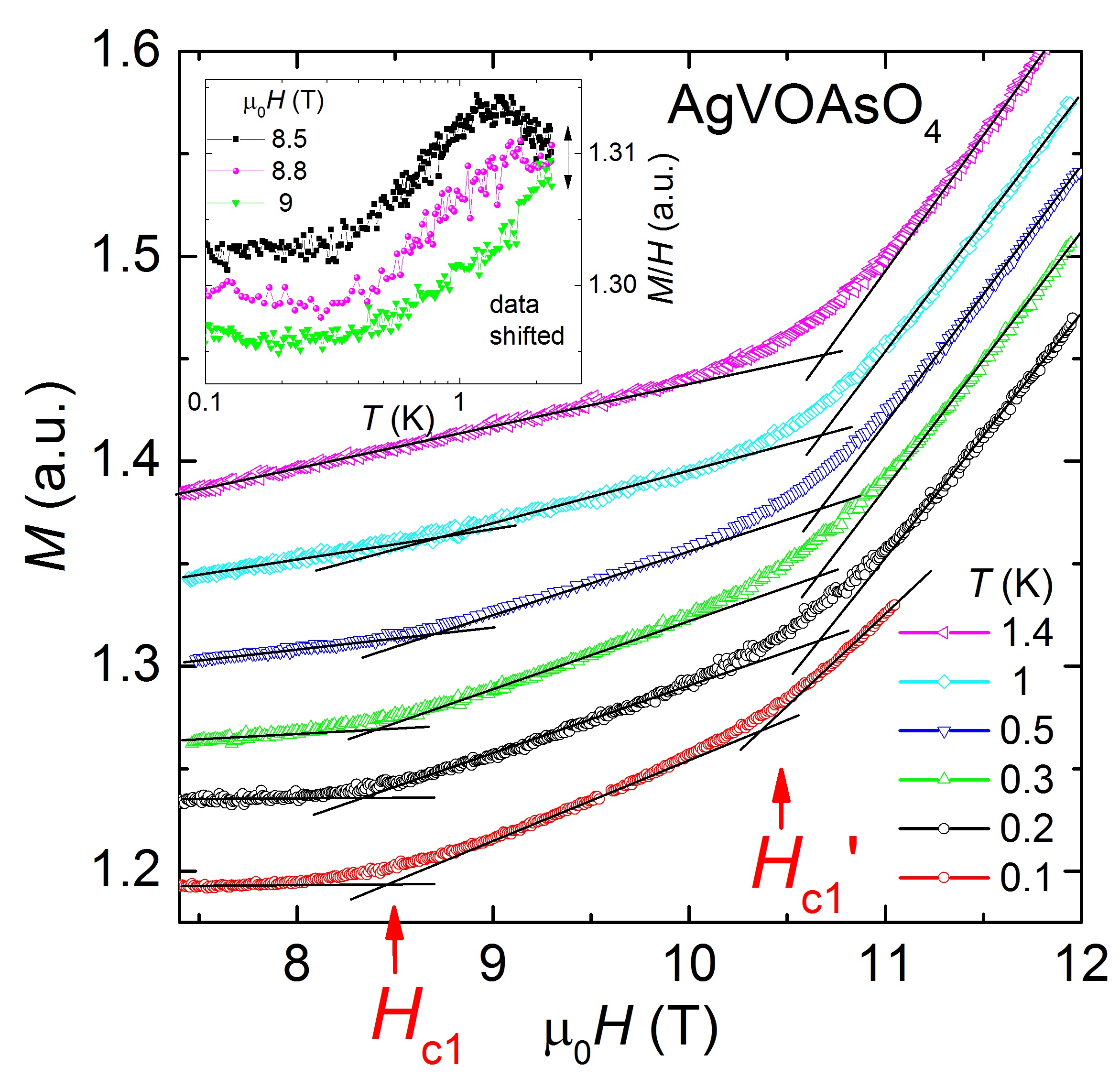}
\caption{\label{MvsH} Magnetization $M(H)$ versus magnetic field $H$ in the field range 7.4~T to 12~T for $T$ between 0.1~K and 1.4~K. The measurements are offset from one another to ensure better visibility. We observe two changes of slope in $M(H)$ at $\mu_{0}H_{c1}$~$=$~8.4~T and $\mu_{0}H_{c1'}$~$=$~10.5~T, which identify two field-induced phase transitions. The inset shows $M(T)/H$ vs $T$ collected in fields of 8.5~T, 8.8~T and 9~T. A broad maximum is observed in this data that is used to define the $H_{c1}$ phase boundary.}
\end{figure}

In order to identify multiple phase transitions in magnetization measurements close to the lower critical field $H_{c1}$, we extend our experiments to lower temperatures. This approach is necessary because quantum fluctuations significantly alter thermodynamic signatures at second order phase transitions close to $T$~$=$~0 in quantum magnets. The effect depends strongly on the mass $m$ of the bosons, which scales with the ratio of the critical fields: $m$~$\propto$~$H_{c1}/H_{c2}$~at $H_{c1}$ and $m$~$\propto$~$H_{c2}/H_{c1}$ at $H_{c2}$\cite{kohama_11}. Asymmetric thermodynamic anomalies (i.e. washed out at $H_{c1}$ and sharp at $H_{c2}$), typical for second order phase transitions, are therefore expected in \Ag. A dilution refrigerator equipped with a 12~T superconducting magnet and a Faraday magnetometer was used to obtain $M(H)$ down to 0.1~K. Fig.~\ref{MvsH} shows $M(H)$ vs $H$ for magnetic fields between 7.4~T and 12~T. We observe two slope changes in $M(H)$, corresponding to two weakly-$T$ dependent phase transitions, at $\mu_{0}H_{c1}$~$=$~8.4~T and at $\mu_{0}H_{c1'}$~$=$~10.5~T. We also carried out $T$-dependent $M(T)$ measurements in constant magnetic fields between 7~T and 12~T and we find a very small $T$-dependence in the data as expected from the light boson mass. A maximum in $M(T)$ is observed at the lower $H_{c1}(T)$ phase boundary, as seen in the inset of Fig.~\ref{MvsH}, that moves to higher $T$ with increasing field, but no clear signature in $M(T)$ is resolved close to the $H_{c1'}(T)$ phase boundary line.

\begin{figure}
\includegraphics[width=3.5in]{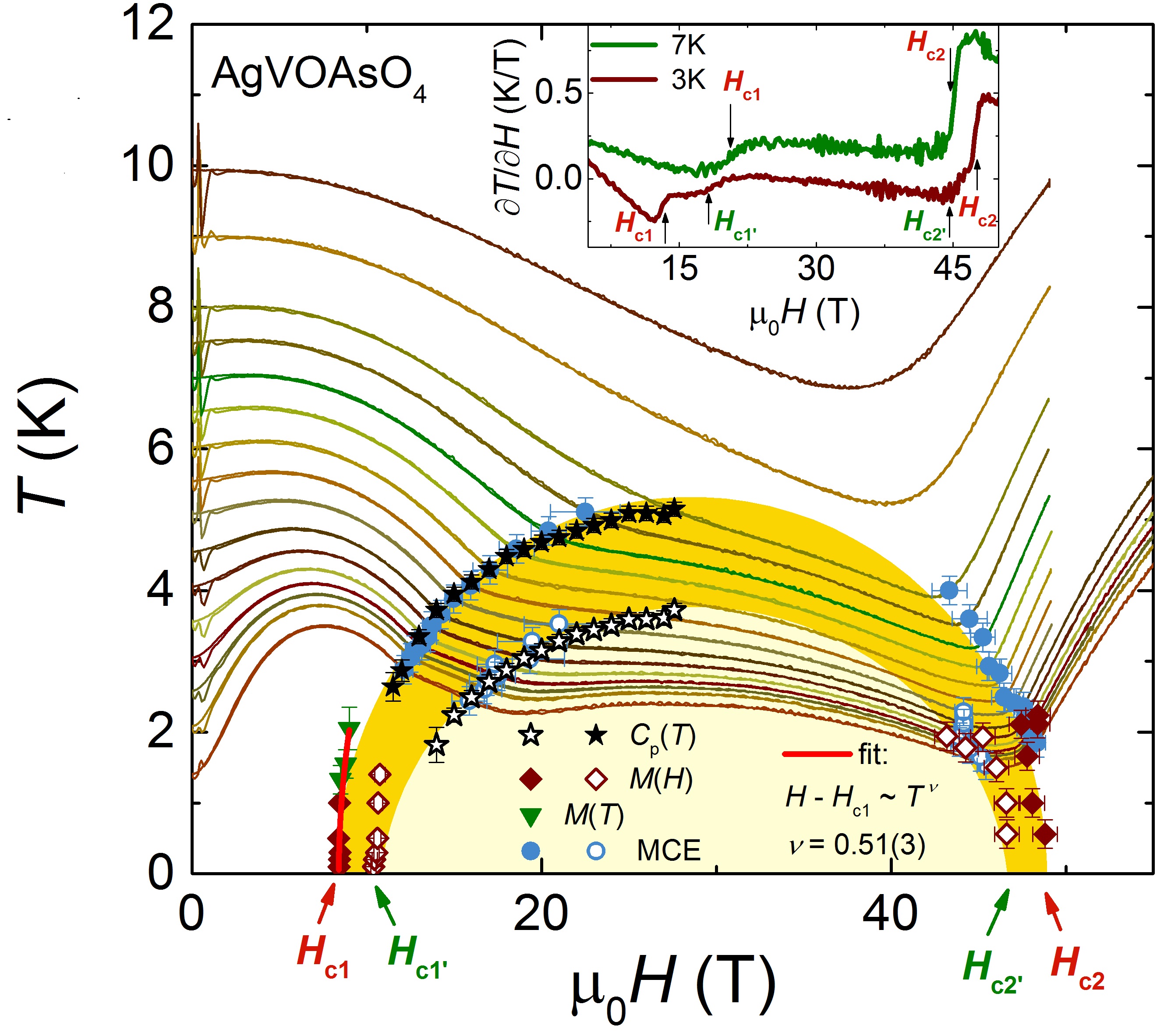}
\caption{\label{phasedia} $H-T$ phase diagram of \Ag obtained from specific heat, MCE, and magnetization experiments. We observe clear signatures of two different field-induced ordered states. Lines of constant entropy obtained under adiabatic conditions in pulsed fields illustrate the MCE and confirm the double dome structure of the phase diagram. A power law fit $H-H_{c1} \propto T^{\nu}$ of the phase boundary at $H_{c1}$ reveals 0.51(3) as the critical exponent. The inset shows the derivative $\partial T/\partial H$ for MCE 
	measurements with zero field temperatures $T_{0}$~$=$~3~K and 7~K. Arrows mark the critical fields $H_{c1}$, $H_{c1'}$, $H_{c2'}$ and $H_{c2}$.}
\end{figure} 

Fig.~\ref{phasedia} summarizes the phase transitions obtained by specific heat and magnetization measurements in pulsed and static magnetic fields. These measurements map out a double-dome phase diagram with $\mu_{0}H_{c1}$~$=$~8.4~T, $\mu_{0}H_{c1'}$~$=$~10.5~T, $\mu_{0}H_{c2'}$~$=$~ 46.6~T and $\mu_{0}H_{c2}$~$=$~48.9~T. We complete our investigations of the high-field properties in \Ag with measurements of the MCE carried out in a capacitor-driven pulsed magnet (pulse duration 36ms) at the International MegaGauss Science Laboratory of the University of Tokyo. In the experiment, the sample is kept under quasi-adiabatic conditions during fast changing field pulses and the sample temperature is monitored, which yields isentropic temperature lines as displayed in Fig.~\ref{phasedia}. Changes in the sample temperature indicate changes of the entropy landscape. For example, the polarization of free spins up to 5~T is reflected in a smooth increase of the temperature because the entropy decreases gradually. A careful analysis of the derivative $\partial T/\partial H$, shown for the $T_{0}$~$=$~3~K and $T_{0}$~$=$~7~K measurements in the inset of Fig.~\ref{phasedia}, reveals four clear steps corresponding to the crossing points of the double phase transitions in the 3K data, whereas only two steps are resolved in the 7K measurement. These results are consistent with the phase lines extracted from specific heat and magnetization measurements. Furthermore, the MCE isentropes exhibit a clear asymmetric behavior with a shallow minimum in $T(H)$ close to $H_{c1}$ indicating a (relative) small entropy accumulation and a deeper minimum around $H_{c2}$ indicating more accumulated entropy. This effect can be explained again with a small boson mass close to $H_{c1}$ and a higher boson mass close to $H_{c2}$. Moreover, the MCE measurements allow us to refine the actual sample temperature in the magnetization experiments when crossing through the phase boundaries because the magnetization and MCE measurements were collected under similar thermodynamic conditions. 

We wish to emphasize that a clear double anomaly observed in both the low and high field regimes of the $H-T$ phase diagram for \Ag cannot be explained by magnetic anisotropy. For a polycrystalline, anisotropic sample, only one broad anomaly in $C(T)/T$ is expected because all grains are randomly-oriented and therefore should contribute equally to the anomaly through a continuous spectrum of critical fields. Also, we found no evidence for anisotropy-induced splitting of the triplet excitation in our INS measurements. Finally, ESR measurements reveal a nearly isotropic g-factor\cite{tsirlin_11}, with $g_{\parallel}$~$=$~1.92 and $g_{\perp}$~$=$~1.96. The 2\% $g$-factor anisotropy is equivalent to a 0.2~T difference in the critical fields $\mu_{0}|H_{c1}-H_{c1'}|$ and less than a 1~T difference for $\mu_{0}|H_{c2}-H_{c2'}|$. These values are much smaller than the 2~T difference in the critical fields that we have identified here on both the low and high field sides of the phase diagram. 

As mentioned before, both phase transitions can be classified as $2^{nd}$ order, because i) the anomaly in $C(T)/T$ shows the typical $\lambda$-shape, ii) no hysteresis or dissipative behavior is observed in the measurements, and iii) we observe asymmetry in the thermodynamic anomalies measured at $H_{c1}$ and $H_{c2}$, which are caused by quantum fluctuations only present at $2^{nd}$ order transitions. Subsequently, we analyze the critical behavior of the phase boundary up to 2~K (Fig.~\ref{phasedia}) with a power law $T \propto H-H_{c1}^{\nu}$ using $\mu_{0}H_{c1}$~$=$~8.4~T and we obtain $\nu$~$=$~0.51$\pm0.13$. This value corroborates the 3D BEC scenario (2/3).

Theoretical work based on density matrix renormalization group (DMRG) calculations for $\alpha$~$=$~0.45 in the alternating chain model finds a complex $H-T$ phase diagram with a dome of incommensurate magnetic order inside a larger dome of commensurate order\cite{maeshima_04,maeshima_05,maeshima_06} when a frustrated coupling exceeding 10\% of the nearest-neighbor intrachain interactions is introduced. The incommensurate ordered phase forms an asymmetric dome and the overall size of this dome increases with increasing frustration. For more than 15\% frustration, the incommensurate state is expected to exceed the phase space corresponding to the commensurate order. Although the main coupling ratio in \Ag is larger than considered theoretically, a similar mechanism may be operative in our case.   


In summary, we have used inelastic neutron scattering on polycrystalline samples of the quantum paramagnet \Ag to confirm that this system is well described by an alternating spin chain model. We have established the $H-T$ phase diagram for \Ag with specific heat, MCE and magnetization measurements in high magnetic fields. We find evidence for a double-dome phase diagram with field-induced order between $\mu_{0}H_{c1}$~$=$~8.4~T and $\mu_{0}H_{c2}$~$=$~48.9~T. A comparison with existing theoretical work for the alternating chain model suggests that the outer phase consists of commensurate order, while the order within the inner dome may be incommensurate. This complex phase diagram establishes \Ag as a promising multi-$\vec{Q}$ BEC candidate capable of hosting exotic topological spin structures. Future NMR or neutron diffraction measurements on single crystals above the lower critical fields $H_{c1}$ and $H_{c1'}$ are essential for elucidating the microscopic spin arrangements of the two field-induced ordered phases in this material.

\begin{acknowledgments}
A portion of this research used resources at the Spallation Neutron Source, a DOE Office of Science User Facility operated by Oak Ridge National Laboratory(ORNL). A portion of this work was performed at the National High Magnetic Field Laboratory, which is supported by National Science Foundation Cooperative Agreement No. DMR-1157490,  the State of Florida and the United States Department of Energy. N. Harrison acknowledges support from the DOE BES project: Science in 100 Tesla.
\end{acknowledgments}

\bibliography{AgVOAsO4_paper}

\end{document}